\documentclass[useAMS,usenatbib]{mn2e}
\usepackage{subfig}
\usepackage{graphicx}
\usepackage{myaasmacros}
\usepackage{times}

\newcommand{\AC}[1]{#1}

\title[Particle tagging]{Particle tagging and its implications for stellar population dynamics}
\author[Le Bret et al.]{Theo Le Bret$^{1,2}$\thanks{E-mail:
theo.lebret@astro.ox.ac.uk (TLB)}, Andrew Pontzen$^{1}$, Andrew P. Cooper$^{3}$, Carlos Frenk$^{3}$, Adi Zolotov$^{4}$, \and 
Alyson M. Brooks$^{5}$, Fabio Governato$^{6}$, Owen H. Parry$^{7}$\\
$^{1}$University College London, London WC1E 6BT , UK\\
$^{2}$Rudolf Peierls Centre for Theoretical Physics, University of Oxford, Oxford, OX1 3NP, UK \\
$^{3}$Institute for Computational Cosmology, Durham University, South Road, Durham, DH1 3LE, UK\\
$^{4}$Center for Cosmology and Astroparticle Physics, Department of Physics, The Ohio State University, OH 43210, USA\\
$^{5}$Rutgers, the State University of New Jersey, Department of Physics \& Astronomy, 136 Frelinghuysen Rd, Piscataway, NJ 08854, USA\\
$^{6}$Astronomy Department, University of Washington, Box 351580, Seattle, WA, 98195-1580\\
$^{7}$Department of Astronomy, University of Maryland, College Park, MD 20742, USA//}
\begin{document}

\date{Accepted     Received }

\pagerange{\pageref{firstpage}--\pageref{lastpage}} \pubyear{2015}

\maketitle

\label{firstpage}

\begin{abstract}

We establish a controlled comparison between the properties of galactic stellar halos obtained with
hydrodynamical simulations and with ‘particle tagging’. Tagging is a fast way to obtain stellar population dynamics: instead of tracking gas and star formation, it `paints'  stars directly onto a suitably defined subset of dark matter particles in a collisionless, dark-matter-only simulation.Our study shows that there are conditions under which particle tagging generates good fits to
the hydrodynamical stellar density profiles of a central Milky-Way-like galaxy and its most
prominent substructure. Phase-space diffusion processes are crucial to
reshaping the distribution of stars in infalling spheroidal systems and hence the final stellar halo.
We conclude that the success of any particular tagging scheme hinges on this diffusion being taken into account, at a minimum by making use of `live' tagging schemes, in which
particles are regularly tagged throughout the evolution of a galaxy.

\end{abstract}

\begin{keywords}
Galactic halo -- simulations: SPH.
\end{keywords}

\section{Introduction} 

Observations of the stellar halo around local group galaxies provide
strong constraints on different models of galaxy formation
\AC{\citep[for historical context see, for instance,][]{eggen62,
    searle78}}.  In today's standard {$\Lambda$}CDM cosmology, large
galaxies such as the Milky Way are predicted to form in part through
the progressive mergers of \AC{smaller} progenitor galaxies
\citep{white78, frenk85, white91, kauffman93}, and the observed stellar halo is
formed, in one way or another, from the remnants of this hierarchical
assembly \citep{helmi99, helmi99b, abadi2006}.

While the current consensus is that \AC{the stellar halo} forms
primarily via accretion of stars from tidally stripped satellites
\AC{\citep[e.g.][]{bell08}} there remain a number of open questions:
for instance, does the halo form exclusively from accretion, or are
some halo stars formed within the main galaxy (referred to as \emph{in
  situ} stars) during dissipative collapse? Are further stars kicked
up from the disc to form an important fraction of the halo
\citep{zolotov09, zolotov10, font2011, tissera13}? Do the main
progenitors of the stellar halo survive as satellite galaxies to the
present day \citep{cooper2010} or are they mostly fully disrupted
\citep{bullock2005}?

In principle, cosmological simulations of galaxy formation which
self-consistently follow the evolution of a baryonic component can be
used to predict the properties of galactic haloes and provide answers
to these questions. In practice, however, this is challenging because
large particle numbers are required to resolve any detail within the
faint, diffuse stellar halo, which contains only a few percent of all
the stars in the galaxy. Only recently has it become possible to
start studying some halo properties in this self-consistent way
\AC{\citep[e.g.][]{zolotov09}}.

As an alternative, some authors have proposed making use of dark
matter (DM)-only N-body simulations --  which are significantly less computationally expensive than hydrodynamical simulations -- where stellar populations are `painted' onto DM
particles to reproduce the collisionless assembly of the stellar halo
\citep{bullock2005, cooper2010, rashkov12}.  Such methods have been
used to make quantitative predictions of halo substructure and
dynamics \citep{cooper11, helmi11, cooper13a, gomez13}, but the
validity of the assumptions underpinning these DM-only models remains
controversial \AC{\citep[e.g.][]{bailin14}}.

So at present there are two techniques for investigating stellar halo
structure -- hydrodynamic simulations and particle tagging -- that
have their own strengths and weaknesses. Recent work has demonstrated
that significant discrepancies between the two approaches can arise
\citep{bailin14}.  In this paper, we will investigate a specific
tagging scheme (that of \citeauthor{cooper2010} 2010) and by comparing
its predictions to those of hydrodynamical simulations, aim to
understand better how and why differences between predictions arise.
While this does not immediately resolve the question of how to produce
``correct" predictions for the stellar halo, it does give a physical
basis for understanding discrepancies and so highlights some essential
prerequisites for realistic modelling.

This paper is structured as follows. In Section 2, we provide an
overview of the tagging technique, stating its main assumptions; in
Section 3, we describe the simulations used in our study, and how they
are used to establish a controlled comparison between tagging and SPH;
in Section 4, we present the outcome of the simulations, and compare
the structure of the different stellar haloes obtained; in Section 5,
we discuss the role of diffusion processes in shaping realistic tagged
haloes, and give a new interpretation for the model-to-model
discrepancies found in the literature. Finally, in Section 6, we
explain the overall physical picture that emerges and highlight how
this should affect the future direction of tagging and broader
investigations of the stellar halo.

\section[methods]{Particle Tagging}

Various methods have been proposed for associating a stellar component
with particles from DM-only simulations \citep{bullock2005,
  cooper2010, libeskind11, rashkov12, bailin14}. DM particles that are
tightly bound to their haloes are assigned `stellar masses' according
to an assumed star formation rate, and the evolution of these
painted particles is then traced up to z = 0, where their final
distribution is taken to \AC{represent} the distribution of stars in
the real stellar halo.

These methods rely on three  assumptions:
\begin{enumerate}
\item Stars form tightly bound to their parent halos (e.g. with
  binding energies typically larger than 90 -- 99\% of the DM
  particles), with energy distributions similar to those of the
  `most-bound' fraction of DM particles. 
\item Recently formed star particles and their tagged DM analogues
  \AC{subsequently follow} similar phase space trajectories. In other
  words, selecting the correct \AC{initial} binding energy for the tagged DM particles is assumed
  to be sufficient to ensure the correct subsequent kinematics. This
  assumption could fail given that, for instance, DM particles and
  stars with similar binding energy will not necessarily have similar
  angular momentum.  
\item Baryonic effects are not important in shaping the stellar halo. 
  For instance, the tidal disruption of subhalos is assumed to be
  unaffected by any interaction between baryons and dark matter and
  the presence of a baryonic disc is typically assumed
  not to affect significantly the distribution of stars in the halo
  \AC{\citep[although some tagging schemes do include effects from an
    analytic disc potential, e.g.][]{bullock2005}}.
  \end{enumerate}

  The first assumption is not likely to be problematic, but the second
  will fail if the stars are supposed to lie in a
  rotationally-supported disc; consequently only the accreted stellar
  halo is usually considered in tagging schemes.  The last assumption
  could affect the validity of tagging in several ways. Firstly, if in
  reality stars kicked up from the disc form a significant part of the
  stellar halo \AC{\citep[see for instance][]{zolotov09, font2011}},
  then tagging cannot be accurate, especially considering also the
  limitation (ii). Secondly, baryonic discs can cause real haloes to
  be less prolate than the tagged DM haloes \citep{bailin14}. Finally,
  the presence of a disc or of \AC{gas expulsion by} SN feedback could
  also affect the dynamics of the DM component of a hydrodynamic
  simulation -- for instance creating cores in satellites
  \citep{pontzen2014}.  Dynamical core creation affects the orbit of
  stars just as it does dark matter
  \citep{navarro96,read05,pontzen2012,maxwell12} and, furthermore, the
  cored satellites are then more easily disrupted
  \AC{\citep{penarrubia2010,parry2012,zolotov12a}}. Conversely,
  baryonic dissipation could give rise to a contraction of the host
  subhalo making it more resilient to tidal disruption.

  The specific tagging scheme we investigate in this paper is
  described by \citet{cooper2010}\AC{, hereafter C10}, where a fixed
  most-bound fraction ($f_{mb}$, usually chosen to lie between 1 and
  10\%) of DM particles in each halo are tagged at every simulation
  output time, \AC{being} assigned `stellar masses' (as well as ages
  and metallicities) according to the prescriptions of a semi-analytic
  model of galaxy formation, GALFORM \AC{(see \citealt{cole2000} for
    details and \citealt{baugh2006} for an overview of hierarchical
    galaxy formation with semi-analytic models)}.

\section[simulations]{Simulations}

\begin{figure*}
    \includegraphics[width=14cm]{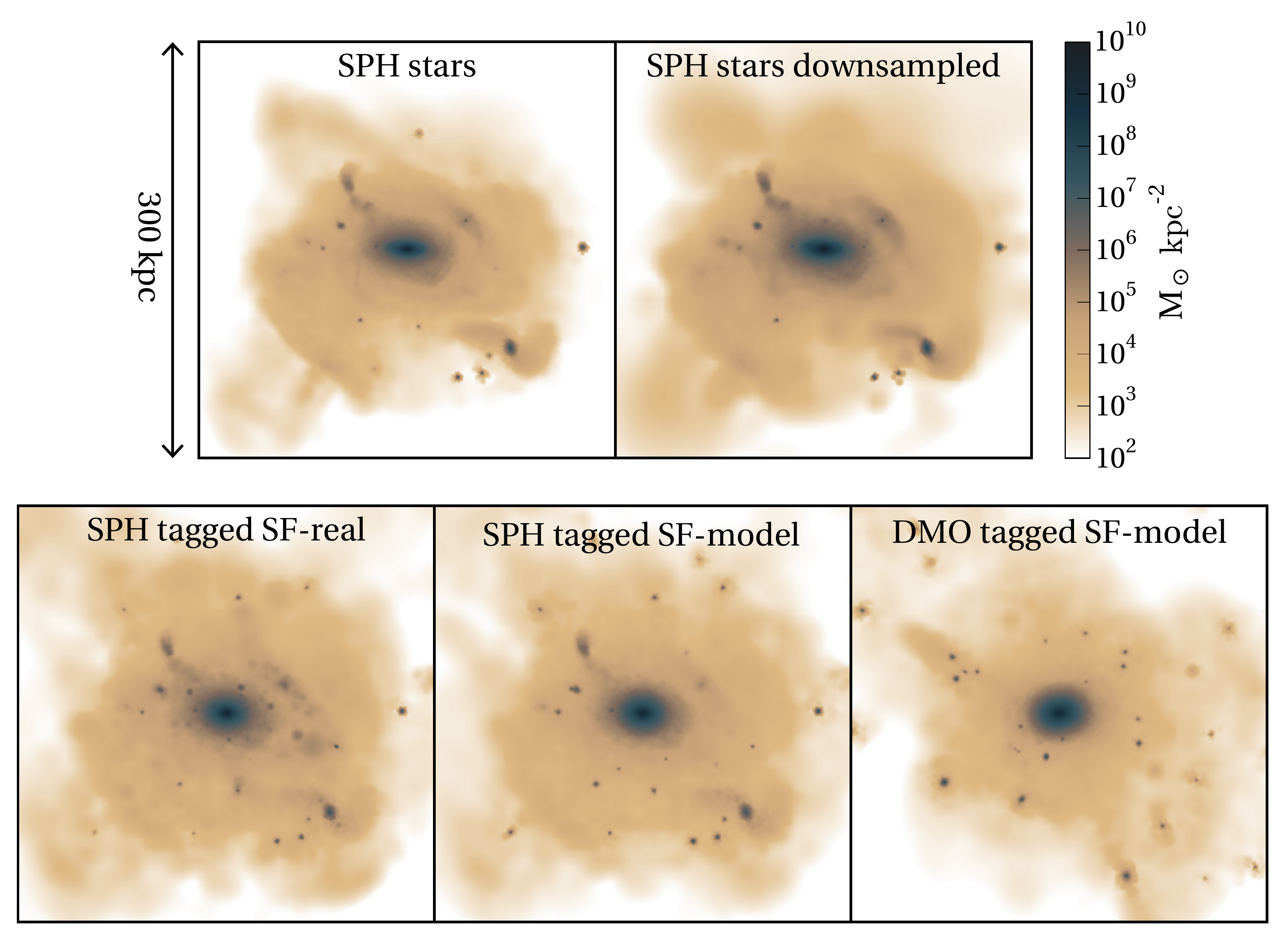}
  \caption{\AC{Projected stellar mass surface density} at \AC{$z=0$} for SPH stars and the `particle tagging' \AC{outputs} for the Seattle simulations. In these panels, tagging is always applied to the top five percent most bound particles (\AC{$f_{\mathrm{mb}=5\%}$}). Note that a one-to-one correspondence between the DMO tagged and SPH tagged is not expected, as small differences between the SPH and DMO runs, combined with the stochastic nature of halo formation, will make them differ qualitatively (see Section~\ref{sec:results}).}
  \label{fig:compareoutputs}
\end{figure*}

We have made use of two sets of simulations, which we call the `Durham' and `Seattle' simulations. Both sets contain smooth-particle-hydrodynamics (SPH) and DM-only (DMO)  simulations of a Milky-Way like galaxy and its immediate environment. They use the volume renormalization or `zoom-in' technique \AC{\citep{katz93}} to increase computational efficiency, with the mass and spatial resolution of the simulation decreasing with radius from the central galaxy. Each adopts a $\Lambda$CDM cosmology; the specific parameters adopted differ slightly with the Seattle (Durham) values being $\Omega_{\rm m}=0.24$ ($0.25$), $\Omega_{\Lambda}=0.76$ ($0.75$),
$\Omega_{\rm b}=0.042$ ($0.045$), $\sigma_{8}=0.77$ ($0.9$), $n_{\rm s}=0.96$ ($1.0$) and in both cases $H_{0}=100h \, {\rm km\,s}^{-1} \, {\rm Mpc}^{-1} = 73 \, {\rm km\,s}^{-1} \, {\rm Mpc}^{-1}$. 
The setup of the two simulations is broadly comparable but crucial differences appear in the handling of star formation feedback as we will discuss below.

The `Seattle' galaxy, sometimes known as h258, was first run by \citet{zolotov09} and then by \citet{zolotov12a} at the higher resolution employed in our current work. Its properties have been extensively discussed in the context of the \cite{BK11} `too-big-to-fail' problem \citep{zolotov12a}; it has also been used as a case study of how baryonic discs can regrow following a major merger at $z=1$ \citep{governato09}. At the end of the simulation it has built a relatively low mass Milky Way ($M_{200}=9\times 10^{11} M_{\odot}$, where $M_{200}$ is the mass enclosed in a sphere with 200 times the mean cosmic density) with a resolution of $1.3 \times 10^5 M_{\odot}$ (DM) and $2.7 \times 10^4 M_{\odot}$ (gas) and force softening of $170$ physical parsecs. 

This simulation was run using \texttt{GASOLINE} \citep{wadsley2004}, a Tree-SPH code. When hydrodynamics is switched on, it uses a pressure-entropy interpolated kernel to eliminate artificial surface tension. Cooling, molecular hydrogen and star formation physics are described in \citet{christensen2012}; feedback is implemented according to the \citet{stinson2006} blastwave model. Self-bound substructures are identified using Amiga's Halo Finder \citep{knollman2009}, hereafter AHF, although we processed the final snapshot with \texttt{SUBFIND} (see below) to ensure the same structures are identified by both algorithms.

The Durham simulation is based on an early version of the PM-Tree-SPH code \texttt{Gadget-3}, and takes the same inital conditions as halo Aq-C of the Aquarius suite \citep{Springel2008}. Its highest particle mass resolution (in a $\sim5 h^{-1}$Mpc region around the target halo) is similar to that of the `level 4' simulation set in Aquarius.  The final mass is $M_{200}=1.8 \times 10^{12} M_{\odot}$, with a resolution of $2.6 \times 10^5\,M_{\odot}$ (DM) and $5.8 \times 10^4 M_{\odot}$ (gas). The force softening is $260$ physical parsecs.

Durham baryonic processes are modelled as described in \citet{Okamoto2010a}, with a number of modifications designed to improve the treatment of supernovae-driven winds, which are explained in detail in \citet{parry2012}. Additional details of this simulation are also presented in \citet{Okamoto2013}. The central galaxy contains a massive centrifugally-supported disc as well as a dispersion-supported spheroid.  Self-bound substructures were identified with the \texttt{SUBFIND} algorithm \citep{Springel01} as modified by \citet{dolag09} to take into account the internal energy of gas particles when computing particle binding energies.

The central galaxy in the Durham simulation (of a $\sim5 h^{-1}$Mpc
region around the target halo) contains a massive
centrifugally-supported disc as well as a dispersion-supported
spheroid.  The final mass of its halo is $M_{200}=1.8 \times 10^{12}
M_{\odot}$. This simulation was used by \citet{parry2012} to study the
satellite system of a Milky Way-like system.  It provided a good match
to the average of the satellite luminosity functions of the MW and M31
but formed its brightest satellites in excessively massive halos,
anticipating what became known as the too-big-to-fail problem
\cite{BK11}. One of the satellites in this simulation generated a core
in its halo as a result of a large inflow of gas into the centre
triggered by a triple merger and its subsequent violent expulsion in a
starburst.

As described above there are a number of numerical and physical
differences between the two sets of simulations. We will see below
that the most important distinction is the type of stellar feedback
implemented. In the Durham case, feedback should be seen as `passive'
in the sense of \cite{pontzen2014}; it produces little or no coupling
between the baryonic and DM components of the simulation. Conversely
the Seattle simulations have `active' feedback which demonstrably
couples baryons and DM with energy being passed from the former to the
latter \citep{pontzen2012}. This means that, for instance, some of the
main satellites in the Seattle simulations develop central density
cores \citep{zolotov12a}, which is not generally the case in the
Durham simulation, with the exception of the single satellite
mentioned above which, in any case, was disrupted shortly after being
accreted into the main halo.  Thus, when comparing haloes and tagged
DM populations in these two sets, we can appreciate the effect of the
very different feedback recipes implemented.

\subsection{Post-processing}

\begin{figure*}
  \begin{center}
    \subfloat{\includegraphics[width=8.5cm, clip, trim = 10mm 0mm 5mm 0mm]{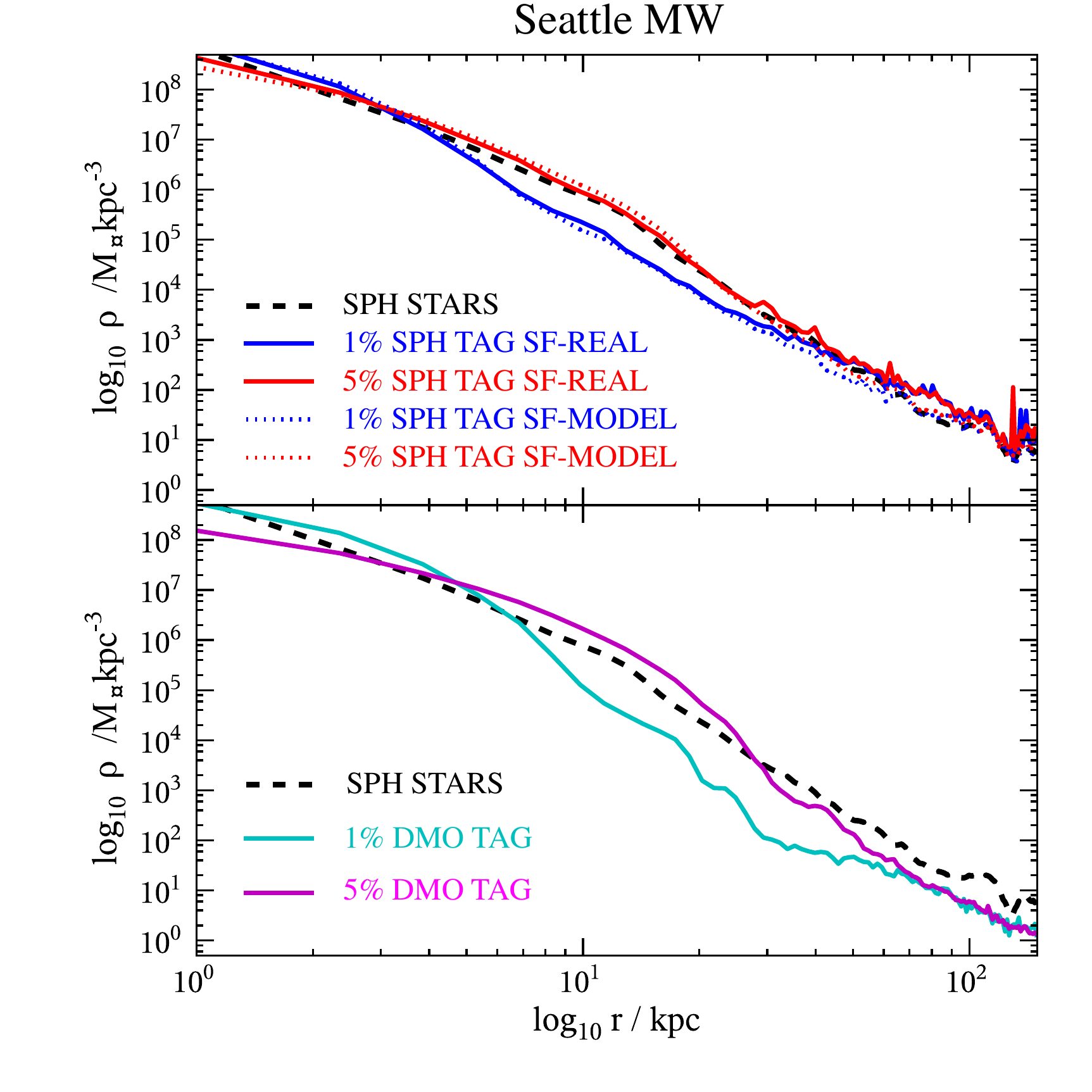}}
    \subfloat{\includegraphics[width=8.5cm, clip, trim = 10mm 0mm 5mm 0mm]{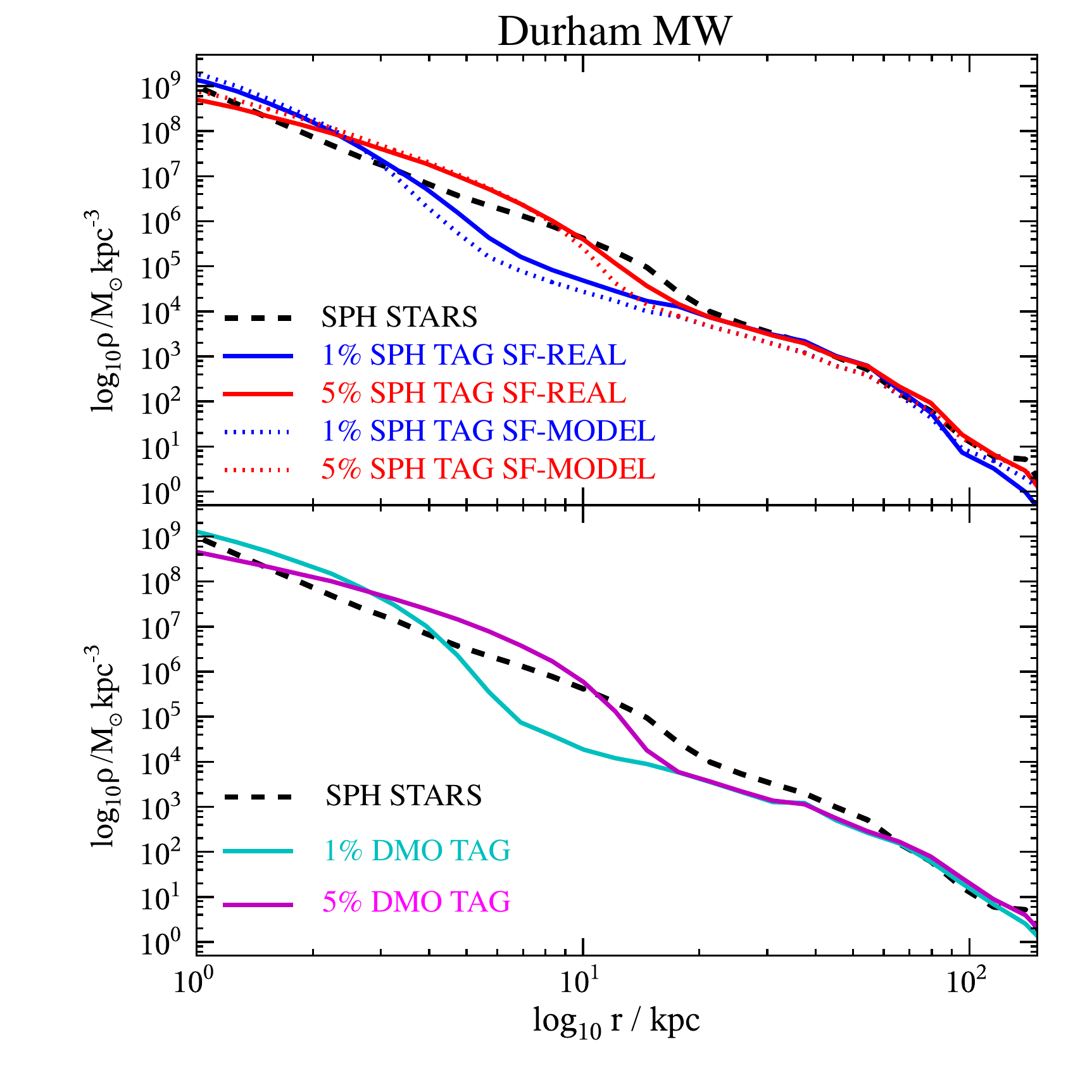}}
  \end{center}
  \caption{Spherically averaged density profile of the main halo, excluding \AC{self-bound substructures}, for the SPH stars and tagged DM, for most bound fractions ($f_{mb}$s) of one and five per cent, in the Seattle (left panel) and Durham (right panel) simulations. The five per cent cases produce output densities within a factor of a few of the SPH results over seven orders of magnitude in stellar density in all cases.}
  \label{fig:profiles}
\end{figure*}

In this paper, we will establish a controlled comparison between the self-consistently formed stars in the simulations described above and ``tagged'' stars in dark-matter-only (DMO) runs of the same regions. For full insight we will also need to study stellar haloes obtained through  algorithms intermediate between these two cases. Our series of approaches are similar to those used by \cite{bailin14}, but differ fundamentally in that they all employ `live' star formation -- associating stellar mass with dark matter particles in a time-dependent way. They are labelled as follows:
\begin{itemize}
\item \texttt{SPH STARS} -- the stellar component of the SPH simulations. These represent an obvious basis from which to assess the success of a tagging scheme, although one should bear in mind that there are a number of differences between SPH and real galaxies; see for instance \cite{stinson10,creasey11}. 
\item \texttt{SPH TAGGED SF-REAL} -- DM particles are tagged within the SPH simulation, and the stellar masses assigned to the particles are calculated using the \AC{actual} star formation rates (SFR) of the SPH galaxies. Thus the star particles are not used explicitly but SFRs are guaranteed to match and baryonic effects such as feedback and the presence of a disc can still affect the tagged DM's dynamics.
\item \texttt{SPH TAGGED SF-MODEL} -- as above, DM particles are tagged in the SPH simulation; but SFRs are calculated using an analytic prescription. Specifically, we assume a power law relation between halo DM mass and \AC{SFR. T}his relation is allowed to vary with redshift, and the power law indices are obtained by fitting to the SPH SFR as a function of halo mass. Note that C10's tagging scheme actually uses the semi-analytic model GALFORM to obtain \AC{SFRs}, but in this paper we adopt a simpler prescription to obtain SFRs consistent with, but no longer identical to, those of the SPH comparison simulations.
\item \texttt{DMO TAGGED SF-MODEL} --  particles are tagged in a DM-only (DMO) run of the galaxy. Ultimately, it is the validity of this approach that we wish to investigate, since it resembles most closely the scheme used by C10 to produce observable predictions of halo properties. The SFRs are calculated in the same way as the \texttt{SPH TAGGED SF-MODEL} run. Note that again we do not use the C10 model to obtain the SFRs since we wish to test the physics of tagging, not of semi-analytic galaxy formation models.  
\end{itemize}

If tagging fails due to an intrinsic difference in the kinematics of DM and star particles, then this should be clear when comparing the haloes obtained in the \texttt{SPH STARS} and \texttt{SPH TAGGED SF-REAL}. If, on the other hand, it fails because baryonic effects do affect significantly the assembly of the stellar halo, then the disagreements should arise when comparing the \texttt{DMO TAGGED SF-MODEL} to the \texttt{SPH STARS} and \texttt{SPH TAGGED SF-REAL} runs. Finally, the \texttt{SPH TAGGED SF-MODEL} is a control run to check that any disagreements between the SPH and DM-only tagging do not in fact occur because the scaling relation used to prescribe the star formation rates is too naive.

\section[results]{Halo structure}\label{sec:results}

\begin{figure*}
   \includegraphics[width=16cm]{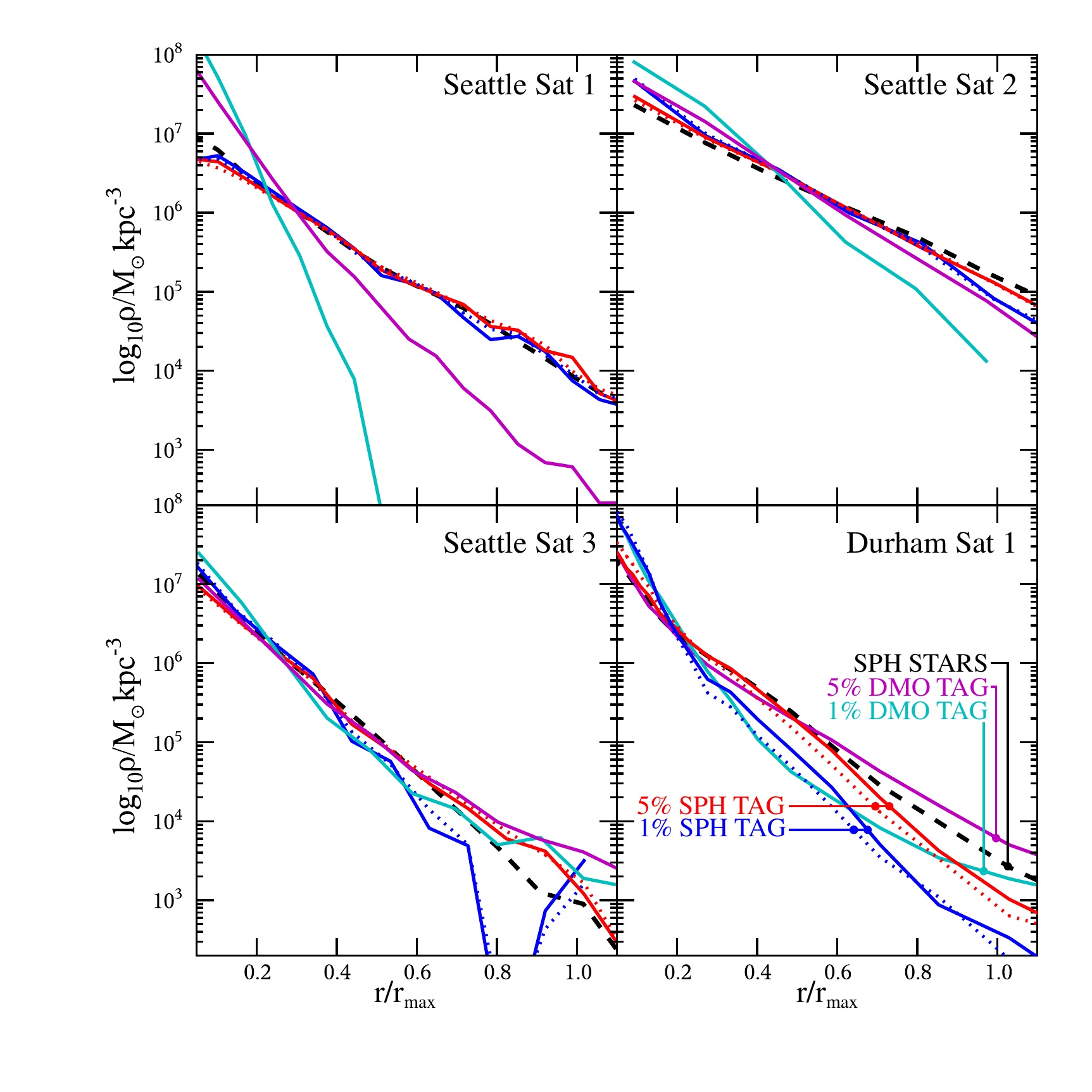}
 \caption{Spherically averaged density profiles of the four most massive satellites for the SPH stars and the different tagging realizations. Densities are normalized so that the total tagged mass equals the SPH stellar mass. Radii are normalized to $r_{\rm max}$, the radius at which the circular velocity achieves its maximum value. The bottom right panel shows the profile of the largest satellite in the Durham simulations. All other satellites come from the Seattle simulations. For Seattle Satellite 1, the scale lengths are significantly underestimated in all DMO cases because the feedback coupling -- that causes collisionless particles including stars to migrate outwards -- is missing. In the other cases reasonable agreement can be achieved using 5\% tagging.}
  \label{fig:satellite_profiles}
\end{figure*}

The principal aim of particle tagging is to produce tagged stellar haloes with realistic \AC{substructures} at \AC{$z = 0$ (such as stellar streams) and having} densities and dynamics comparable to those of real haloes.  The purpose of the present paper is to investigate some of the physics relevant to stellar halo formation rather than to tackle  observational matters -- this will be left to a companion paper, Cooper et al (in prep) -- but we nonetheless need to start with a brief comparison of the observable outputs.

To obtain a qualitative view of how well DM tagging reproduces the features of the SPH \AC{stellar halo, we} obtained 2-D projected density maps of the SPH stars and their tagged DM analogues at \AC{$z = 0$}, tagging first the DM in the SPH, and then in the DM-only simulation (Figure \ref{fig:compareoutputs}).  
We find that the \texttt{SPH TAGGED} outputs produce a stellar halo containing much the same \AC{substructure} as the SPH stars: in both cases, stellar streams are clearly discernible, associated with tidally stripped satellites of the main galaxy, and seem to have comparable densities. This suggests that the SPH tagged DM and SPH stars have similar density distributions by the time the simulations reach \AC{$z = 0$} -- though this does not necessarily imply that the SPH stars and tagged DM were in agreement at the time of tagging\AC{. T}his is an important distinction which will be discussed further in the next section. 

The resolution in the tagged DM tends to be lower than in the SPH stars. This is not an intrinsic limitation of particle tagging, but an artefact of the way we carry out the comparison: here we tag DM in an SPH simulation (and a corresponding DM-only simulation), where the mass and \AC{spatial} resolution in the DM component is lower by a factor of about 10 than those achieved by recent pure \AC{$N$}-body codes (such as the Aquarius haloes, on which C10 applied their tagging scheme).  In order to compare the tagged and SPH star images more fairly, one can thus downsample the latter, taking only every fifth particle (top right panel of Figure \ref{fig:compareoutputs}). The effect of this is to decrease the resolution of the streams, which is also observed in the \texttt{SPH TAGGED}. In other words, the fact that substructure is somewhat `blurred'  in the tagging images is related to the artificially low resolution.  

In fact the resolution issues become even more subtle and can be seen to be responsible for the visible tendency of tagging to produce larger abundances of \AC{subhaloes}, even in the \texttt{SPH TAGGED SF-REAL} case where the effects of baryonic feedback on SFRs are directly taken \AC{into} account, and where tidal interactions with the galactic disc still have the ability to disrupt satellite haloes. The number of dark matter particles tagged is independent of the star formation rate; therefore, in a halo with low SFR, a large number of particles may  be tagged with very small stellar masses. In this situation the mass resolution locally becomes much higher than that of the SPH stars, which have fixed mass per particle at birth irrespective of SFR. 
So, although overall the resolution is lower in the tagged halo, we verified that for small subhaloes the resolution \AC{is} in fact larger in the \AC{tagging realizations} than in the SPH stars. 

There seems to be fairly little qualitative difference between the \texttt{SPH TAGGED SF-REAL} and \texttt{SPH TAGGED SF-MODEL} runs, which suggests that the simple prescription used for assigning SFRs to DM haloes is adequate for our purposes.
In the case of the DMO tagged simulation a qualitative comparison is less straightforward, as small differences between the SPH and the DMO simulation runs, combined with the chaotic nature of halo structure formation, means that even the locations of DM substructures are quite different. \AC{Although streams and substructures are  present,} they do not correspond directly to the SPH streams, and there seem to be a significantly larger number of \AC{subhaloes} than in the SPH case. We will argue below that this last difference \AC{may,} at least in some 
\AC{cases,} be an effect of the `bursty' feedback implemented in the Seattle simulations, where some \AC{satellites develop} DM cores in their inner regions, and thus become more susceptible to tidal disruption when passing through the galactic disc -- see \citet{zolotov12a, zolotov14} and \AC{also \citet{penarrubia2010}}.

For a more quantitative comparison between the SPH stars and tagged haloes, we turn to spherically-averaged density profiles for the main halo and a few satellite galaxies.
Figure \ref{fig:profiles} shows the main stellar halo density for various tagging realizations for the SPH and DMO cases (top and bottom panels) in the Seattle and Durham simulations (left and right). In each panel, the thick black dashed line shows the \texttt{SPH STARS}. For all three tagging realizations, taking $f_{mb}=5\%$ produces output densities within a factor of a few of the SPH results over seven orders of magnitude in stellar density. Notably, the Seattle \texttt{SPH TAGGED} cases are in particularly good agreement -- this result can be tied to strong active star formation feedback and will be discussed further below and in Section \ref{sec:diffusion}.

Conversely the worst agreement arises with \texttt{DMO} tagging within the inner few kpc of the main galaxy. Tagging will be a poor description of the star distribution in regions where baryons dominate, \AC{i.e.} near the disc and bulge of the main galaxy. The inaccuracies can then propagate outwards if a significant fraction of the halo stars were originally disc stars that were scattered out to the stellar halo. A more detailed discussion of this process and its importance in Seattle simulations can be found in \citet{zolotov09}.

We also obtained density profiles for the main satellites (Figure \ref{fig:satellite_profiles}). In reading order, the figure shows the first to third most massive satellite in the Seattle simulations, then the most massive of the Durham satellites. Once again the SPH stars are shown by the thick dashed line -- one can read off that the agreement between the SPH stars and tagging $f_{mb}=5\%$ in both SPH and DMO is broadly quite good. However the most massive Seattle satellite (top left panel) shows a striking discrepancy between the \texttt{DMO} and \texttt{SPH} cases. It has previously been shown for the Seattle simulations that the SN feedback causes the dynamics of the DM and baryonic component to couple strongly, creating DM density cores \citep{pontzen2012} in the central regions. Tagged and `real' star particles alike will be thrown to larger radii by this process. But in the DMO simulation, the process is of course missing -- so the \texttt{DMO TAGGED} cases form stellar components with significantly too short a scale length. This begins to show that the agreement between tagging and SPH will depend on what kind of feedback recipe is used in the SPH (i.e. `active' \AC{vs.} `passive').

\begin{figure}
    \includegraphics[width=8.5cm]{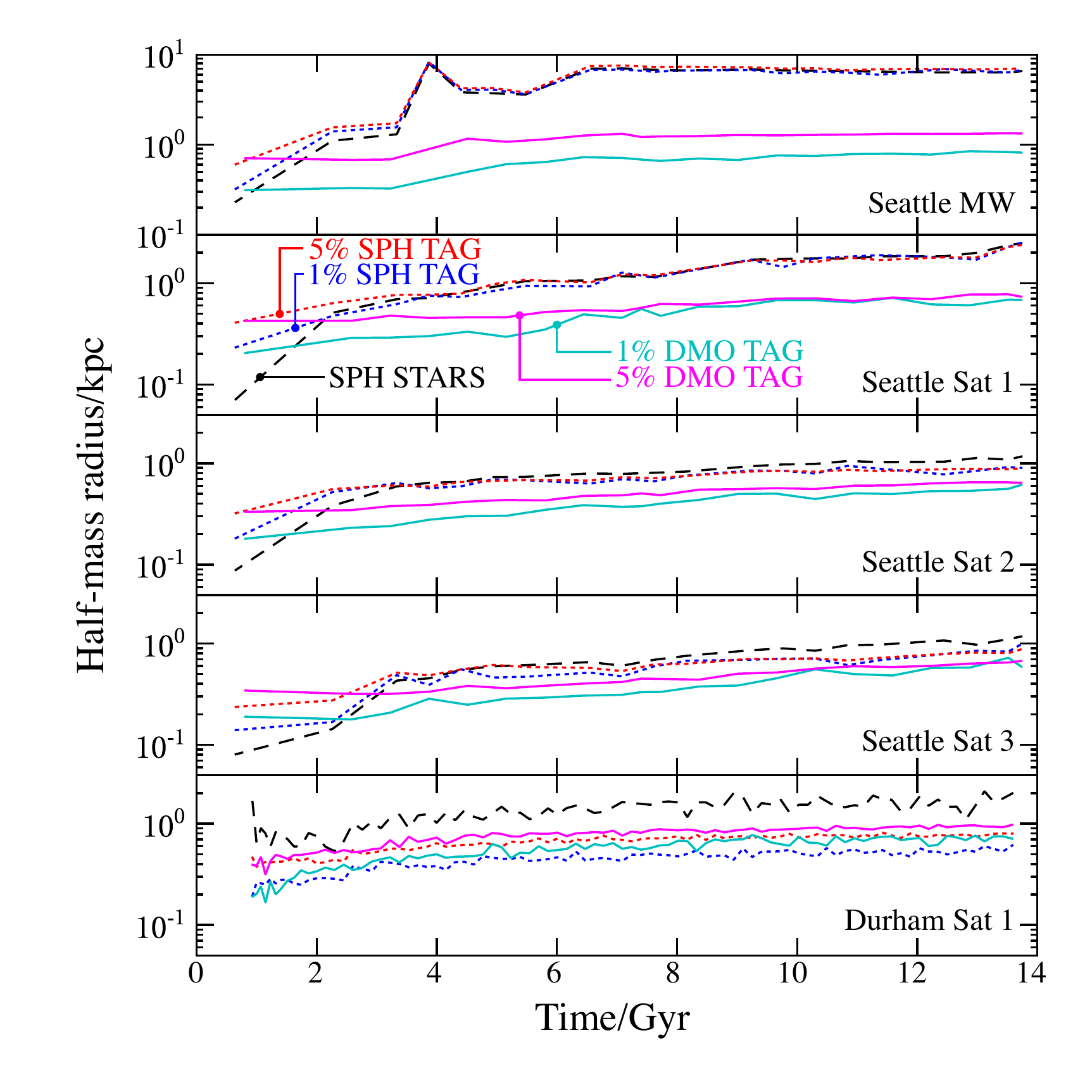}
  \caption{Evolution of the half-mass radius of different tagged populations and the stars they are meant to represent in the Seattle main galaxy (top panel) and its three largest satellites (next 3 panels, in order of satellite size). The lowest panel shows Durham satellite 1. The colour scheme is as for the density profiles: black for the SPH stars, blue and red for the SPH tagged 1 and tagged 5 respectively, and cyan and magenta for the DM-only tagged 1 and tagged 5 respectively. In all cases, the half-mass radius at early times is sensitive to the tagging mode, but at late times this dependence is substantially reduced. This is tied to diffusion of particle energies as described in the text, and is a particularly strong effect when `active' feedback is present as in the Seattle SPH simulations. }
  \label{fig:r50}
\end{figure}

In fact, this deep connection between the mode of stellar feedback and
the behaviour of tagging algorithms can be further reinforced by
studying the difference between the different tagging fractions
($f_{mb}=$1\% or 5\%, hereafter ``tag~1'' and ``tag~5''). 
As we commented briefly above, in the Seattle \texttt{SPH TAGGED} outputs setting \AC{1\% or 5\%} seems to make surprisingly little difference to the final density profiles of the halo, especially in the satellites, or in the main galaxy in the outer regions. Increasing the number of tagged particles by a factor of 5 does not for instance change the scale length of the exponential density profiles, or their general shape. This suggests that the overall density profiles are primarily being set by dynamical processes {\it after tagging}, not by the details of the tagging itself. The expected increased prominence of these processes when feedback is active is reflected in near-complete insensitivity to $f_{mb}$ in the Seattle \texttt{SPH TAGGED} cases.

\begin{figure*}
  \begin{center}
   \includegraphics[width=\textwidth]{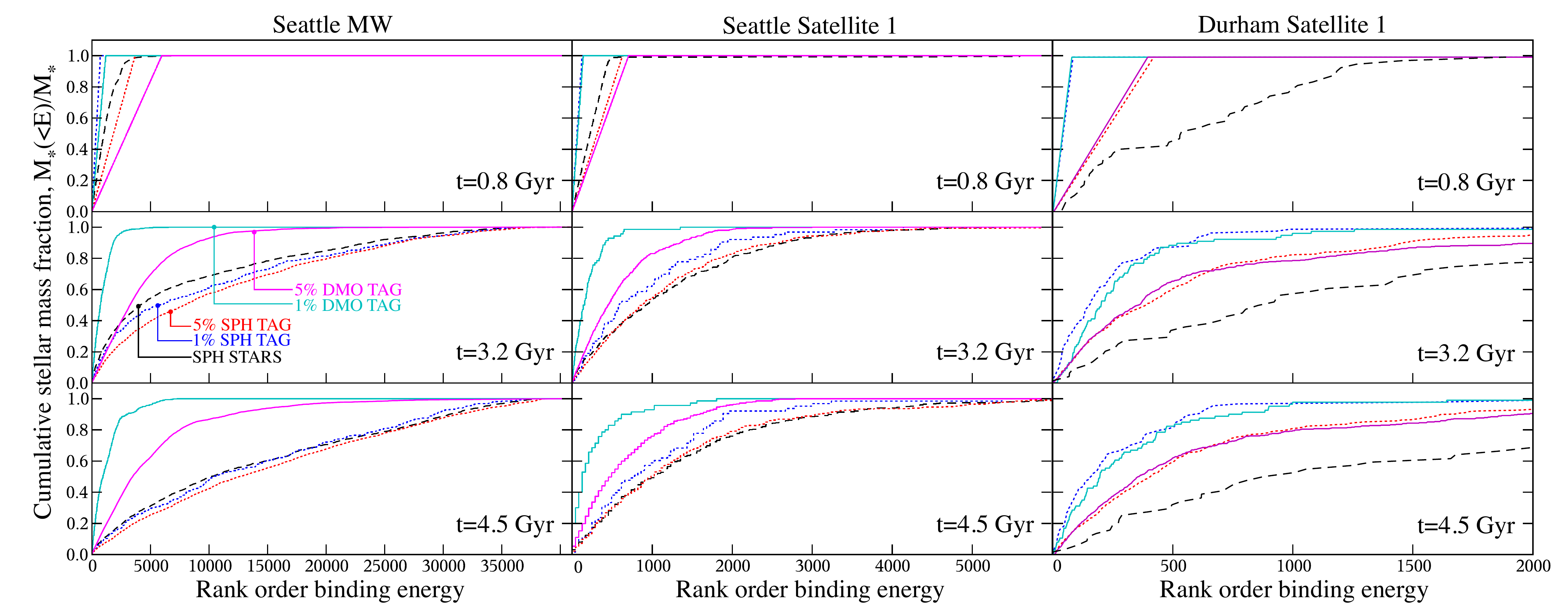}    
  \end{center}
  \caption{Diffusion in the rank binding energy distribution of recently formed stars and their analogue tagged DM in the Seattle main halo (left panel) and largest satellite (middle panel), as well as the Durham largest satellite (right panel). From top to bottom, the panels represent the simulations at simulation output times: t = 800 Myr, t = 3200, t = 4500, where the stars are those formed at t = 800 Myr, and the DM has been assigned a stellar mass at the same time. The evolution over time is characterised by a drift to larger scale radii, and (in most cases) towards closer agreement between initially distinct populations. As also seen in Figure \protect\ref{fig:r50}, the effect is strongest in the Seattle SPH cases where active feedback generates strong potential fluctuations.}
  \label{fig:diffusion}
\end{figure*}

The Durham simulations have a more passive type of feedback and are therefore missing the dynamic redistribution of stars that we are claiming.  The bottom right panel of Figure \ref{fig:satellite_profiles} shows the profile of the largest satellite in the Durham simulations: agreement between the tagging realizations and the SPH stars is not as good as for the Seattle satellites. Moreover, the SPH tag 1 and tag 5 runs do not resemble one another as closely as they do in the Seattle satellites. This is suggestive, but we now turn to a more explicit way to verify that diffusion of stars through the energy space is a significant effect.

%
% DIFFUSION SECTION
%

\section[diffusion]{Diffusion}\label{sec:diffusion}

In the previous section we  compared the stellar haloes from our various tagging realizations and found, from a number of perspectives,  agreement between apparently very different approaches. We have suggested that these can be tied to dynamical redistribution of the stars after they have been formed; we will now provide a direct analysis to support that claim.

We first trace the evolution of the radial scale length of a population of tagged particles over a period of a few Gyr after its initial assignment, as follows. In an early simulation snapshot ($z = 8$), we select the most-bound 1 and 5 percent of DM particles, as well as the recently formed stars. We then obtain the radius enclosing half of the total mass of these particles. We track these same particles over time, calculating their half-mass radius at each snapshot. The results are shown in Figure \ref{fig:r50} for (top to bottom panels) the major progenitor and the three most massive satellites in the Seattle simulations, followed by the most massive satellite in the Durham simulations.

At the tagging snapshot, $z=8$ (far left of each panel), the half-mass radii do not typically match those of the stars, with initial discrepancies between \texttt{SPH TAGGED SF-MODEL} (dotted lines), \texttt{DM TAGGED SF-MODEL} (solid lines) and \texttt{SPH STARS} (dashed line) reaching an order of magnitude. The difference between tagging the 1 or 5 percent most-bound DM particles in the halos is also  clearly discernable: as naively expected, when a larger number (by a factor of 5) of particles are tagged, their distribution is more diffuse, and the half-mass radius is larger.

However, as we follow the evolution of this distribution, all populations ``diffuse" outwards to larger radii; initially compact populations tend to diffuse faster and so the initial differences in scale radii are significantly eroded.  Particularly in the Seattle cases, the tagged particles of the \texttt{SPH} runs (dotted lines) tend to converge to mimic the stars (dashed line) --  irrespective of the binding energies at which they were originally selected. In these cases the convergence between the half-mass radii of the tag 1, tag 5 and stars occurs on a timescale of around 3~Gyr, and these three initially distinct populations become virtually indistinguishable. 

In the DM-only tagged assignments (solid lines) convergence is slower, but still significant. In Seattle satellites 2 and 3, for example, the half-mass radius discrepancy narrows from an initial discrepancy of a factor 2 at $z=8$ to agreement by $z=0$. However, unlike in the SPH case, it takes the entire Hubble time for this to occur and in the case of satellite 2 it does not even run to completion. Furthermore in the central galaxy the diffusion process ceases after a major merger at a time of around 4~Gyr. This is consistent with a picture of diffusion being driven primarily by time-dependence in the potential of a spheroidal halo -- after this time the galaxy is quiescent and, in the case of the SPH run, has developed a stable disc structure.

For each satellite, the diffusion is faster and more sustained in the SPH case than in the DMO case. This follows because baryonic feedback enhances the process by producing more stochastic time-dependence in the potential, particularly in the case of satellite 1 (second panel from top in Figure \ref{fig:r50}) where the populations from the SPH simulation continue migrating outwards throughout cosmic time. As discussed in Section \ref{sec:results}, this satellite produces a feedback-driven dark matter density core in the hydrodynamic cases. The outward migration is a symptom of this process and by definition is not present in the DMO cases. 

With the above in mind, we can start to see the diffusion process as a double-edged sword. On the one hand it does tend to drive tagging realizations into better agreement, because it reduces sensitivity to the initial choice of which particles to tag. On the other, the strength of the diffusion can be enhanced by baryonic processes leading to a new, fundamental source of discrepancy with DMO simulations. In the Durham cases (illustrated here by Satellite 1, the lowest panel of Figure \ref{fig:r50}) the DMO and SPH cases are in much better agreement, but the convergence is also much weaker. Both these aspects follow from the relatively passive approach to feedback taken.

In practice, this means that -- if one believes in the kind of bursty star formation histories that lead to the `active' feedback coupling -- DMO tagging schemes will need to tag particles that are less bound than the regions in which star formation actually takes place, because they miss processes that kick stars up to higher energies. Concretely, the lower left panel of Figure \ref{fig:profiles} shows 5 percent tagging in the Seattle DMO simulations produce better fits to the SPH star density profiles  than 1 percent tagging -- and yet Figure \ref{fig:r50} shows directly that at the time of formation, the SPH STARS are more tightly bound even than the 1 percent case.

\subsection{Energy and angular momentum}

We can obtain another view of the importance of diffusion by studying the process in energy space. First, we obtain the energy distributions of \AC{recently formed} stars and their tagged DM analogues (using different values of $f_{\mathrm{mb}}$) at the time the DM is assigned a stellar mass, and track the evolution of these distributions \AC{(for the same set of particles)} over a few \AC{gigayears}. We confirm that, at the time of tagging, the recent stars and most-bound DM particles do have very different energies (top panels, Figure \ref{fig:diffusion}; the three panels from left to right show Seattle MW, Satellite 1 and Durham Satellite 1). 

One can now see explicitly how in the Seattle SPH cases, the baryonic feedback drives the full \texttt{SPH STARS} and \texttt{SPH TAG MODEL-SF} energy distributions into agreement, irrespective of the starting $f_{\mathrm{mb}}$. Once again the DMO cases do evolve to reduce initially stark differences in the energy distributions, but this diffusion does not complete and the energy distributions of this population of stars are significantly different from those of the SPH case. 

Comparing satellite 1 from the Seattle and Durham SPH simulations (middle and right panel respectively), one can see explicitly how the difference between active and passive feedback is crucial. In the Durham case, the differences in the diffusion between SPH and DMO are much less marked because the gas evolution has relatively little impact on the collisionless dynamics. Once again we see the dual impact of diffusion -- it substantially reduces  sensitivity to the initial distribution of star formation events, but the extent of diffusion depends on the nature of feedback. 

\begin{figure}
   \includegraphics[width=0.5\textwidth]{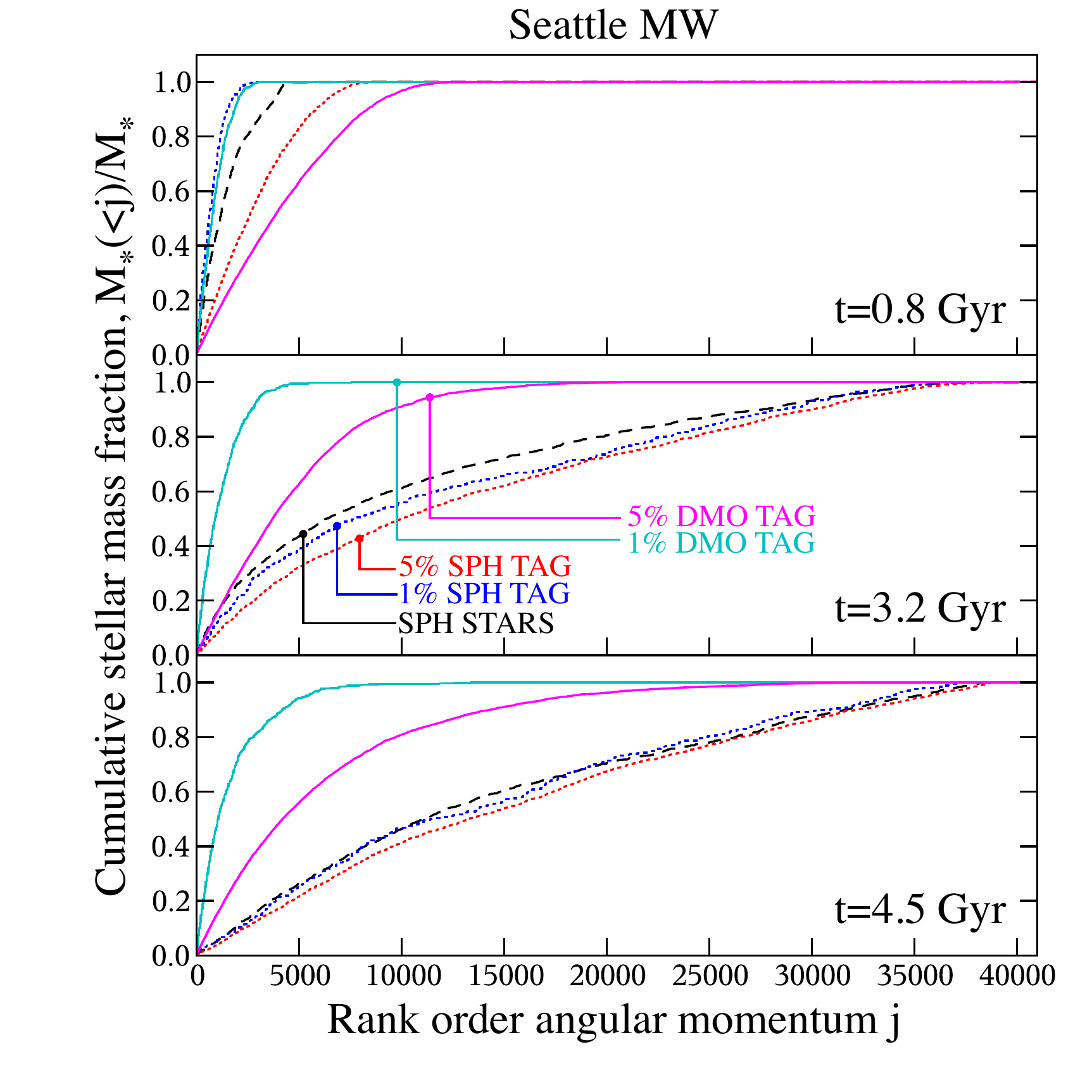}
   \caption{The angular momentum distribution of a population of stars and their analogue tagged DM in the main halo, selected at high redshift (z = 8) in the Seattle simulations. Stars like these that end up in the halo, rather than the disk, have distributions of angular momentum evolves rapidly -- similar to how their energy distribution changes (Figure \ref{fig:diffusion}). Stars forming later and contributing to the disk in the SPH case retain their angular momentum.}  \label{fig:angmom_diff}
\end{figure}

Finally we verified that a very similar view can be obtained in angular momentum space (Figure \ref{fig:angmom_diff}) -- the chaos associated with potential fluctuations from mergers or stellar feedback causes the distributions to drift away from their initial arbitrary shapes. Only when stars form in a stable disc is the diffusion in both angular momentum and energy stopped. 

In conclusion, there is very strong evidence for the diffusion of orbits through phase space being a crucial ingredient in deciding the final distribution of stars in spheroidal systems. This in turn has significant implications for interpreting the existing literature and for future schemes attempting particle tagging, as we now discuss.

\section{Conclusions and discussion}

We have re-implemented and investigated a particle tagging scheme for simulating stellar haloes in DM-only simulations\AC{, first} used and described in \citet{cooper2010}. The scheme assigns a `stellar mass' to the most-bound DM particles in each halo at every simulation output time. In our case, we use this method to establish a comparison similar in spirit to that of \cite{bailin14}, constructing a series of different stellar halos which increasingly differ from the SPH stars. Our focus is on understanding the physics that sets the distribution of stars within each subhalo, and hence in the stellar halo of a Milky Way-like galaxy.

Our main conclusions are:

\begin{itemize}
   \item Tagging a fixed fraction of dark matter within SPH simulations reproduces qualitative features of the SPH stellar halo structure and sub-structure (Figure \ref{fig:compareoutputs}); tagged stellar haloes have density profiles very similar to the SPH haloes (compare red and blue 'tagged' lines to the black dashed line in Figure \ref{fig:profiles}).  
   \item Performing the same experiment in the DMO run produces more pronounced differences between the tagged halo and reference SPH stars halo, although for 5\% tagging one obtains similar scale-lengths and density profiles that deviate from the SPH stars by less than one dex (see Figure \ref{fig:satellite_profiles}). The exception to this pattern is  Seattle's satellite 1, which has a large dark matter core in the SPH run; therefore, by definition, collisionless particles in the DMO run end up on the `wrong' orbits relative to the SPH case.
   \item All the above results can be tied to diffusion of the tagged DM particles through energy space. Populations which are tagged in differing regions nonetheless end up with similar phase space distributions if sufficient time is given between the initial assignment and the point at which the stellar halo properties (densities, satellite profiles and so on) are measured (e.g. Figures \ref{fig:r50} and \ref{fig:diffusion}). 
   \item The agreement between tagging and SPH is affected by the type of SN feedback recipes implemented in the SPH sub-grid physics. The Seattle simulations have a strong, `bursty' feedback recipe \citep{stinson2006}. We have already commented on the core in satellite 1, but even elsewhere in the Seattle simulations we see considerably accelerated diffusion compared to the DMO runs  (e.g. Figure \ref{fig:diffusion}). Conversely the Durham simulations (which have `passive' feedback, rather than Seattle's `active' feedback; \citeauthor{pontzen2014} 2014)  drive the diffusion at rates much more similar to the DMO case.
\end{itemize}

Another way to phrase our conclusions is that (in a spheroidal, pressure-supported structure) once a star has formed, the collisionless dynamics will make  initially different phase space distributions of the tagged and star particles converge. Future work on stellar tagging needs to take this into account. Key recommendations are that (a) tagging must be done on-the-fly rather than at infall (see further discussion below); (b) one should consider tagging DM in regions considerably larger than the strict star formation region of a halo, to make up for missing energy kicks that might arise from astrophysics in the real universe. 

The recent work of \cite{bailin14} focused on comparing observable results between SPH stars and  a somewhat different particle tagging algorithm. The key difference between their work and ours is that their stars in a given satellite are tagged at the time of maximum mass (essentially just prior to infall), rather than dynamically throughout the simulation. The diffusion process therefore operates for a much shorter time in \cite{bailin14} and, accordingly, they see more stark differences between tagged and dynamic stars than we do. 

Diffusion itself may be driven in the DM-only simulation by mergers or tidal interactions \citep{stickley12, kandrup03} and accelerated in the SPH simulation by baryonic processes such as interaction with a disc or SN feedback \citep{valluri13}. Further study of these processes will be necessary to understand more fully the exact domain of applicability of particle tagging.

Other proposals for improving and understanding tagging were given by \citet{bailin14} and focus on selecting a more appropriate region of the phase space, in particular by considering angular momentum. Our work suggests that, in a spheroidal system, differences in initial angular momentum get smoothed out in much the same way as the differences in energy. Therefore selecting the `correct' particles to tag is as much about studying dynamical evolution as it is about characterising the orbits of recently formed stars.

\section*{Acknowledgments}
TLB acknowledges an ERC studentship. AP is supported by a Royal Society University Research Fellowship. APC is supported by a COFUND/Durham Junior Research Fellowship under
EU grant agreement 267209. FG was funded by NSF grants AST-0908499, AST-0607819 and NASA 13-ATP13-0020. CSF acknowledges ERC Advanced
Investigator Grant COSMIWAY and support from Science and Technology
Facilities Council grant ST/F001166/1,ST/L00075X/1. FG was funded by
NSF grants AST-0908499, AST-0607819 and NASA 13-ATP13-0020. OHP was supported by NASA grant NNX10AH10G and by NSF grant CMMI1125285. We made use of pynbody N-body analysis software \citep{pynbody} in our analysis for this paper (https://github.com/pynbody/pynbody). This research used the DiRAC Facility, jointly funded by STFC and the Large Facilities Capital Fund of BIS. The Seattle simulations were run at NASA HEC.

\bibliography{bibliography}

\bsp

\label{lastpage}

\end{document}